\documentclass[letter]{aa} % for the letters 
\usepackage{amsmath}
\usepackage{graphicx}
\usepackage{graphics}
\usepackage{subfigure}
\usepackage{txfonts}
\usepackage{natbib}
\usepackage{xcolor}
\bibpunct{(}{)}{;}{a}{}{,}

\def\teff{\ifmmode T_{\rm eff} \else $T_{\mathrm{eff}}$\fi}

\def\ltsima{$\buildrel<\over\sim$}
\def\lsim{\lower.5ex\hbox{\ltsima}}

\newcommand{\hii}{H~{\sc ii}}
\newcommand{\ha}{\ifmmode {\rm H}\alpha \else H$\alpha$\fi}
\newcommand{\hb}{\ifmmode {\rm H}\beta \else H$\beta$\fi}
\newcommand{\lya}{\ifmmode {\rm Ly}\alpha \else Ly$\alpha$\fi}

\newcommand{\heii}{He~{\sc ii}}
\newcommand{\Heiiuv}{He~{\sc ii} $\lambda$1640}
\newcommand{\Heiiopt}{He~{\sc ii} $\lambda$4686}

\newcommand{\ebv}{\ifmmode E_{\rm B-V} \else $E_{\rm B-V}$\fi}
\newcommand{\av}{\ifmmode A_{\rm V} \else $A_{\rm V}$\fi}
\def\ergscm{erg s$^{-1}$ cm$^{-2}$}
\def\msun{\ifmmode M_{\odot} \else M$_{\odot}$\fi}
\def\msunyr{\ifmmode M_{\odot} {\rm yr}^{-1} \else M$_{\odot}$ yr$^{-1}$\fi}
\def\zsun{\ifmmode Z_{\odot} \else Z$_{\odot}$\fi}

\def\lsun{\ifmmode L_{\odot} \else L$_{\odot}$\fi}

\def\mup{\ifmmode M_{\rm up} \else M$_{\rm up}$\fi}
\def\mlow{\ifmmode M_{\rm low} \else M$_{\rm low}$\fi}
\def\aap{A\&A}

\def\apjl{ApJL}
\def\apjs{ApJS}

\newcommand{\oh}{\ifmmode 12 + \log({\rm O/H}) \else$12 + \log({\rm
O/H})$\fi}
\newcommand{\oiii}{[O~{\sc iii}]}
\def\Oii{[O~{\sc ii}] $\lambda$3727}
\def\Oiii{[O~{\sc iii}] $\lambda\lambda$4959,5007}
\def\oiiil{[O~{\sc iii}]$\lambda 5007$}
\def\flyf{\ifmmode f_{\rm Lyf} \else $f_{\rm Lyf}$\fi}
\def\pz{\ifmmode P(z) \else $P(z)$\fi}
\def\ki2{\ifmmode \chi^2 \else $\chi^2$\fi}
\def\zphot{\ifmmode z_{\rm phot} \else $z_{\rm phot}$\fi}

\newcommand{\xphot}{\ifmmode x_\gamma \else $v_\gamma$\fi}
\newcommand{\xobs}{\ifmmode x_{\rm obs} \else $x_{\rm obs}$\fi}
\newcommand{\xcmf}{\ifmmode x_{\rm CMF} \else $x_{\rm CMF}$\fi}
\newcommand{\vexp}{\ifmmode V_{\rm exp} \else $V_{\rm exp}$\fi}
\newcommand{\vmax}{\ifmmode V_{\rm max} \else $V_{\rm max}$\fi}
\newcommand{\nh}{\ifmmode N_{\rm HI} \else $N_{\rm HI}$\fi}
\newcommand{\dv}{\ifmmode \Delta v({\rm em-abs}) \else $\Delta v({\rm em}-{\rm abs})$\fi}

\def\fesc{\ifmmode f_{\rm esc} \else $f_{\rm esc}$\fi}

\def\frellya{\ifmmode f^{\rm rel}_{\rm{Ly}\alpha} \else $f^{\rm rel}_{\rm{Ly}\alpha}$\fi}

\def\hii{H{\sc ii}}
\newcommand{\mstar}{\ifmmode M_\star \else $M_\star$\fi}
\newcommand{\muv}{\ifmmode M_{1500} \else $M_{1500}$\fi}
\newcommand{\auv}{\ifmmode A_{\rm UV} \else $A_{\rm UV}$\fi}
\newcommand{\luv}{\ifmmode L_{\rm UV} \else $L_{\rm UV}$\fi}
\newcommand{\lir}{\ifmmode L_{\rm IR} \else $L_{\rm IR}$\fi}
\newcommand{\lbol}{\ifmmode L_{\rm bol} \else $L_{\rm bol}$\fi}
\newcommand{\liruv}{\ifmmode L_{\rm IR+UV} \else $L_{\rm IR+UV}$\fi}
\newcommand{\liroveruv}{\ifmmode L_{\rm IR}/L_{\rm UV} \else $L_{\rm IR}/L_{\rm UV}$\fi}
\newcommand{\nlyc}{\ifmmode N_{\rm Lyc} \else $N_{\rm Lyc} $\fi}
\newcommand{\rholyc}{\ifmmode \rho_{\rm Lyc} \else $\rho_{\rm Lyc} $\fi}
\newcommand{\chion}{\ifmmode \xi_{\rm ion} \else $\xi_{\rm ion}$\fi}
\newcommand{\chioncorr}{\ifmmode \xi_{\rm ion}^0 \else $\xi_{\rm ion}^0$\fi}

\newcommand{\Civ}{C~{\sc iv} $\lambda$1550}
\newcommand{\Ciii}{C~{\sc iii}]}
\newcommand{\Ciiiuv}{C~{\sc iii}] $\lambda$1909}
\newcommand{\Oiiiuv}{O~{\sc iii}] $\lambda$1666}
\newcommand{\source}{J1154+2443}
\newcommand{\paper}{Izotov2018J11542443:-a-lo}

\begin{document}

%
%    \title{Efficiency of compact $z \sim 0.3$ Lyman continuum leakers of producing ionizing photons
%and comparison with
%    high-redshift galaxies}
    \title{Intense C~{\sc iii}] $\lambda\lambda$1907,1909 emission from a  strong Lyman continuum emitting galaxy}
  \subtitle{}
  \author{D. Schaerer\inst{1,2}, 
Y. I. Izotov$^{3}$,
K. Nakajima$^{4}$, 
G. Worseck$^{5}$, 
J. Chisholm$^{1}$, 
A. Verhamme$^{1}$,
T.X. Thuan$^{6}$,
S. de Barros$^{1}$
}
%  \offprints{}
  \institute{Observatoire de Gen\`eve, Universit\'e de Gen\`eve, 51 Ch. des Maillettes, 1290 Versoix, Switzerland
         \and
CNRS, IRAP, 14 Avenue E. Belin, 31400 Toulouse, France
        \and
Bogolyubov Institute for Theoretical Physics,
National Academy of Sciences of Ukraine, 14-b Metrolohichna str., Kyiv,
03143, Ukraine
%Main Astronomical Observatory, Ukrainian National Academy of Sciences,
%27 Zabolotnoho str., Kyiv 03680, Ukraine
        \and
National Astronomical Observatory of Japan, 2-21-1 Osawa, Mitaka,
Tokyo 181-8588, Japan
        \and
Institut f\"ur Physik und Astronomie, Universit\"at Potsdam, Karl-Liebknecht-Str. 24/25, D-14476 Potsdam, Germany
        \and
Astronomy Department, University of Virginia, P.O. Box 400325, 
Charlottesville, VA 22904-4325, USA
         }

\authorrunning{D.\ Schaerer et al.}
\titlerunning{Strong \Ciii\ emission from a strong Lyman continuum emitter}

\date{Received date; accepted date}
%\date{Accepted for publication in A\&A Letters}

%\abstract{CONTEXT}
%{AIMS
%}
%{METHODS
%}
%{RESULTS
%}
%{CONCLUSIONS
%}
%% 5 {} token are mandatory

\abstract{We have obtained the first complete ultraviolet (UV) spectrum of a strong Lyman continuum(LyC) emitter at low redshift --
the compact, low-metallicity, star-forming galaxy \source\ --\ with a Lyman continuum escape fraction of 46\% discovered recently.
% by \cite{\paper}.
The Space Telescope Imaging Spectrograph spectrum shows strong \lya\ and \Ciiiuv\ emission, as well as \Oiiiuv.
Our observations show that strong LyC emitters can have UV emission lines with a high equivalent width (e.g.\ EW(\Ciii)$=11.7 \pm2.9$ \AA\
rest-frame), although their equivalent widths should be reduced due to the loss of ionizing photons.  
%This shows that 
The intrinsic ionizing photon production efficiency of \source\ is high, $\log(\chioncorr)=25.56$ erg$^{-1}$ Hz, comparable to that
of other recently discovered $z \sim 0.3-0.4$ LyC emitters.
Combining our measurements and earlier determinations from the literature, we find a trend of increasing \chioncorr\ with increasing 
\Ciiiuv\ equivalent width, which can be understood by a combination of decreasing stellar population age and metallicity.
Simple ionization and density-bounded photoionization models can explain the main observational features including 
the UV spectrum of \source.
}


 \maketitle

%%%%%%%%%%%%%%%%%%%%%%%%%%%%%%%%%%%%%%%%%%%%%%%%%%%%%%%%%%%%%%%%%%%%%%%%%%%%%%%%%s
\section{Introduction}
\label{s_intro}
To improve our understanding of galaxies, their ISM, and ionizing properties at high redshift, UV spectra play a fundamental
role, especially as this domain is accessible to ground-based telescopes over a very wide redshift range, and observations
of the most distant galaxies become feasible with the largest telescopes.
Therefore both observational works and studies improving and applying the diagnostic power of rest-UV lines have flourished recently
\citep[see eg.][]{Stark2014Ultraviolet-emi,Stark2017Lyalpha-and-C-I,Fevre2017The-VIMOS-Ultra,Maseda2017The-MUSE-Hubble,Jaskot2016Photoionization,Gutkin2016Modelling-the-n,Nakajima2018The-VIMOS-Ultra}

At low redshift, there are  relatively few UV observations of star-forming galaxies covering the spectral range of $1200-2000$ \AA,
which includes for example\ \lya, \Civ,  \Heiiuv, O~{\sc iii}] $\lambda\lambda$1660,1666 (hereafter \Oiiiuv), 
C~{\sc iii}] $\lambda\lambda$1907,1909 (hereafter \Ciiiuv) and other emission lines, although the need 
for comparison samples has recently been recognized \citep[cf.][]{Rigby2015C-III-Emission-,Berg2016Carbon-and-Oxyg,Senchyna2017Ultraviolet-spe}. 
Despite this, only approximately 28 sources with \Ciii\ emission line detections from the HST are currently known from these studies. 
However, and most importantly, none of them is known as a Lyman continuum (LyC) emitter, 
and follow-up observations in the LyC are too time consuming for these $z \ll 0.3$ sources.

Since Lyman continuum emitters are obviously fundamental to understand the sources of cosmic reionization,
it is of prime interest to study such sources in terms of their physical properties, interstellar medium (ISM), stellar populations, and so on.
Building on the recent success in identifying LyC emitters with HST at $z \sim 0.3$
\citep{Leitherer2016Direct-Detectio,Izotov2016Eight-per-cent-,Izotov2016Detection-of-hi,\paper,Izotov2018Low-redshift-Ly},
we have targeted one of the strongest LyC leakers, the compact $z=0.369$ galaxy \source\ with
a LyC escape fraction of 46\% from \cite{\paper} to obtain the first complete HST UV spectrum of a strong LyC emitter.

%%%%%%%%%%%%%%%%%%%%%%%%%%%%%%%%%%%%%%%%%%%%%%%%%%%%%%%%%%%%%%%%%%%%%%%%%%%%%%%%%
\section{The UV spectrum of a compact $z=0.369 $ LyC leaker with a high escape fraction}
\label{s_obs}

\subsection{HST observations}
The strong LyC emitter \source\ was observed 
as part of mid-cycle observations in may 2018 (GO 15433, PI Schaerer). 
The observations were taken with the Space Telescope Imaging Spectrograph (STIS) NUV-MAMA  using
the grating G230L with the central wavelength 2376 \AA\ and the slit 52\arcsec x0\farcs 5
resulting in a spectral resolution $R=750$. 
The acquisition was done using an image with the  MIRVIS filter and an exposure of 
360s. The science exposure was 10418s obtained during four consecutive 
sub-exposures. Data reduction was done using the STIS pipeline. 
The individual sub-exposures with somewhat different spectral ranges were co-added
using the 
%\LEt{Please spell out all acronyms the first time they
%appear in the paper, followed by the abbreviation in parentheses, both in
%the abstract and again in the main text. After that, please only use the
%abbreviation. See A and A language guide Section 5.2.4 www.aanda.org/language-editing}
IRAF dispcor routine, and the one-dimensional (1D) spectrum was extracted using IRAFs splot. 

The observed 
%\LEt{Please spell out all acronyms the first time they
%appear in the paper, followed by the abbreviation in parentheses, both in
%the abstract and again in the main text. After that, please only use the
%abbreviation. See A and A language guide Section 5.2.4 www.aanda.org/language-editing}
STIS spectrum (Fig.\ \ref{fig_spectrum}) shows a clear detection of \Ciiiuv, the presence
of \Oiiiuv, plus strong \lya, which was previously observed by \cite{Izotov2018J11542443:-a-lo}.
The measured line fluxes and equivalent widths are listed in Table \ref{table1}.
We note that with the chosen resolution, the \Ciii\ and O~{\sc iii}] doublets are not resolved.

\setcounter{figure}{0}
\begin{figure}[tb]
{\centering
\includegraphics[width=9cm]{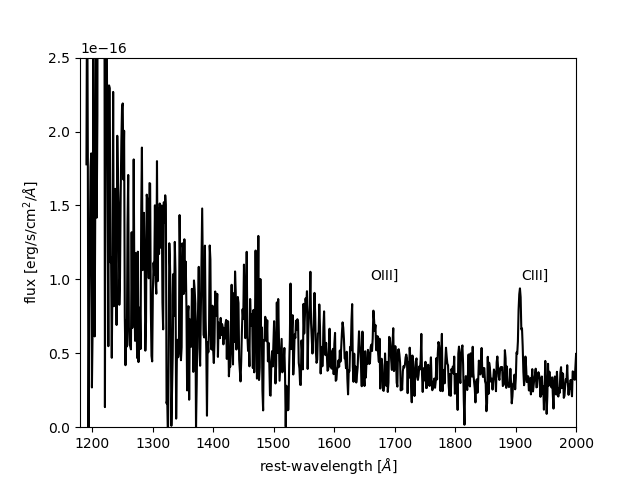}
\caption{STIS spectrum of \source\ showing the detection of the \Oiiiuv\ and \Ciiiuv\ lines.}
}
\label{fig_spectrum}
\end{figure}

\begin{table}[htb]
{\small
%{\scriptsize
\caption{Measurements from the STIS spectrum of \source.}
\begin{center}
\begin{tabular}{lrrrrrrr}
Line & Observed flux$^a$ & Rest-frame equivalent width$^b$ \\
\hline \\ 
\lya\      &    $92.64 \pm 3.80$ &133$^c$  \\
\Heiiuv  &    $<0.68$ ($1 \sigma$) & $<2.9$ \\
\Oiiiuv\  &    $4.34 \pm 0.73$  &    $5.8 \pm  2.9$  \\  % EW observed 8+/- 4 
\Ciiiuv\      &    $6.57 \pm 0.38$   &   $11.7 \pm 2.9$ \\   %                     16+/- 4
\hline
\multicolumn{3}{l}{$^a$ in units of $10^{-16}$ \ergscm; \,\,\, $^b$ in \AA}\\
%\multicolumn{3}{l}{$^b$ in \AA} \\
\multicolumn{3}{l}{$^c$ From \cite{Izotov2018J11542443:-a-lo}}
\end{tabular}
\end{center}
\label{table1}
}
\end{table}

% % % % % % % % % % % % % % % % % % % % % % %
\subsection{\lya\ emission}
Compared to the UV spectrum obtained with the Cosmic Origins Spectrograph (COS) in a circular aperture 
of 2\farcs 5 diameter, the  \lya\ flux observed with the 0\farcs 5
slit of STIS is a factor 1.6 lower, whereas the UV continuum fluxes are 
compatible within the errors \citep[cf.][]{Izotov2018J11542443:-a-lo}.
The total \lya\ equivalent width, EW, of \source\   determined from the COS data is EW(\lya)=133 \AA\ rest-frame.
From the STIS 1D spectrum we find compact \lya\ emission with an extension of 4.2 pix (0\farcs 1 FWHM), 
slightly larger than the UV continuum with a FHWM of 0\farcs 077.

% % % % % % % % % % % % % % % % % % % % % % %
\subsection{The C/O abundance}
The observed \Ciii/\Oiii\ line ratio and the \Oiiiuv\ equivalent width of \source\
are comparable to other star-forming galaxies
\cite[cf.][]{Berg2016Carbon-and-Oxyg,Senchyna2017Ultraviolet-spe}.
Compared to the latter sample, 
the EW(\Oiiiuv)$\sim 6$ \AA\ we find is at the upper end of the values measured 
by \cite{Senchyna2017Ultraviolet-spe} for their sample, which shows EW$\sim 0.7 -7$ \AA.

We derived the C/O abundance from the UV line ratio and the ratio of the UV to optical lines
for the electron temperature and density derived in \cite{\paper}.
The UV line fluxes were corrected for Milky Way and internal attenuation, also following this latter paper.
%using $c(\hb)=0.07$, $R_V=2.4$, and  the \citet{cardelli89} law, as also derived in this paper.
%
% Line intensities were corrected for MW extinction with C(Hb)=0.02 +
% Cardelli law with Rv=3.1 at observed wavelengths, and for internal extinction
%with C(Hb) = 0.07 + Cardelli with Rv=3.1 at rest-frame wavelengths.
%
We obtain $\log({\rm C/O}) = -0.99 \pm 0.16$ using the \Ciiiuv\ and \Oiiiuv\ lines following \cite{Garnett1995The-evolution-o}, 
and $\log({\rm C/O}) = -0.84 \pm 0.06$ from the \Ciii\ and \Oiii\ line ratio \citep{Izotov1999Heavy-Element-A}.
We assumed an  ionization correction factor ${\rm ICF}=1.282$ from \cite{Garnett1995The-evolution-o}. 
Other assumptions affecting the C/O abundances are\ discussed in \cite{Gutkin2016Modelling-the-n}, for example.
Both values are comparable to other C/O observations in \hii\ regions and star-forming galaxies of similar
metallicity \citep[cf.][]{Berg2016Carbon-and-Oxyg,Amorin2017Analogues-of-pr}.

\begin{figure}[tb]
{\centering
\includegraphics[width=9cm]{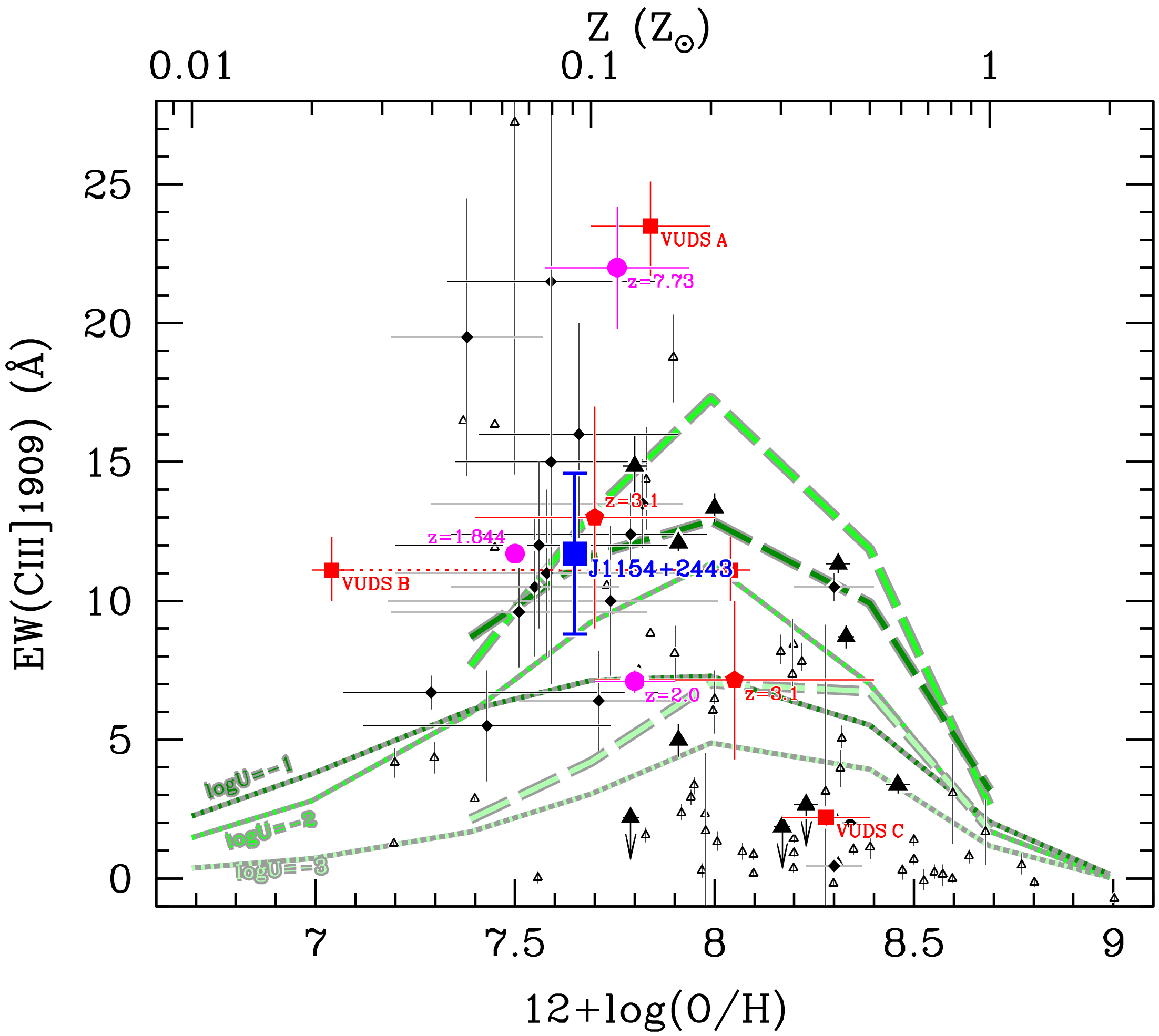}}
\caption{\Ciiiuv\ rest-frame equivalent widths for low- and high-$z$ star-forming galaxies with measured metallicities.
The filled blue square shows \source\ and 
the black filled triangles denote the low-redshift galaxies studied by \cite{Senchyna2017Ultraviolet-spe} and reanalysed by
\cite{Chevallard2018Physical-proper}.
Red symbols show the z=2--4 galaxies (stacked spectra) analysed by \cite{Nakajima2018The-VIMOS-Ultra,Nakajima2018The-Mean-Ultrav} and
magenta symbols show the $z = 1.844$, 2.0, and 7.73 galaxies from  \cite{Berg2018A-Window-On-The},  \cite{Erb2010Physical-Condit},
and \cite{Stark2017Lyalpha-and-C-I}, respectively.
The smaller open triangles and filled diamonds represent the other low-$z$ and
$z=2-4$ galaxies, respectively, compiled in \cite{Nakajima2018The-VIMOS-Ultra}.
Green curves present photoionization models using single- (dotted) and binary-
(long dashed) stellar populations with different ionization parameters \citep{Nakajima2018The-VIMOS-Ultra}.}
\label{fig_c3}
\end{figure}
% % % % % % % % % % % % % % % % % % % % % % %
\subsection{Strong \Ciiiuv\ emission}
The \Ciiiuv\ line of \source\ has a high equivalent width, EW$($\Ciii$) = 11.7 \pm 2.9$ \AA\
rest-frame, comparable to the strongest \Ciii\ emitters at low redshift with metallicities $\oh \sim 7.5-8$,
which are similar to that  of \source\ \citep[$\oh =7.62-7.65$, ][]{Izotov2018J11542443:-a-lo},
and also comparable to high-$z$ galaxies as illustrated in Fig.\ \ref{fig_c3}.
\source\ agrees well with the proposed relation between the \lya\ and \Ciii\ equivalent widths reported
by \citet{Stark2014Ultraviolet-emi}, and with measurements from larger samples, although the latter show
a significant scatter between these observables \citep[cf.][]{Fevre2017The-VIMOS-Ultra}.

As shown in Fig.\ \ref{fig_c3}, it is found empirically that the \Ciii\ equivalent width of star-forming galaxies increases with
decreasing metallicity \citep[cf.][]{Rigby2015C-III-Emission-,Senchyna2017Ultraviolet-spe,Nakajima2018The-VIMOS-Ultra}.
Since the C/O abundance ratio of \source\ is ``normal" for this metallicity
we conclude that the \Ciii\ equivalent width can be compared to that of other sources at the 
same metallicity.
Photoionization models approximately follow the observed trend, although they predict a non-monotonic behaviour
with a maximum EW(\Ciii)  at $\oh \approx 7.8-8$, and a decrease of the equivalent widths at lower metallicities 
as also shown in Fig.\ \ref{fig_c3} \citep[cf.][]{Nakajima2018The-VIMOS-Ultra}.
More data is needed to confirm the predicted decrease of  EW(\Ciii)  at very low metallicities.

The main result of these observations is probably the fact that the  \Ciiiuv\  equivalent width
is very high despite the fact that this galaxy has a very high escape fraction of Lyman continuum
photons, with $\fesc = 0.46  \pm 0.02$ according to \cite{\paper}.
If a significant fraction of Lyman continuum photons escape the galaxy, the \Ciii\ equivalent width 
should be reduced.
Indeed such a behaviour is predicted for high ionization parameters in the models
of \cite{Jaskot2016Photoionization}, where it is seen that EW(\Ciii)  decreases almost linearly with \fesc\
for $\log U \ga -2$.  Despite this, the  EW of \source\ is  high; it is even among the highest observed. 
This shows that the \Ciii\ line cannot reliably be used to select strong Lyman continuum emitters, 
as proposed by \cite{Jaskot2016Photoionization}, who suggest that {\em low} \Ciii\ equivalent widths 
together with other emission lines, such as \Oiii/\Oii, would indicate density bounded cases.

\begin{figure}[t]
{\centering
\includegraphics[width=9cm]{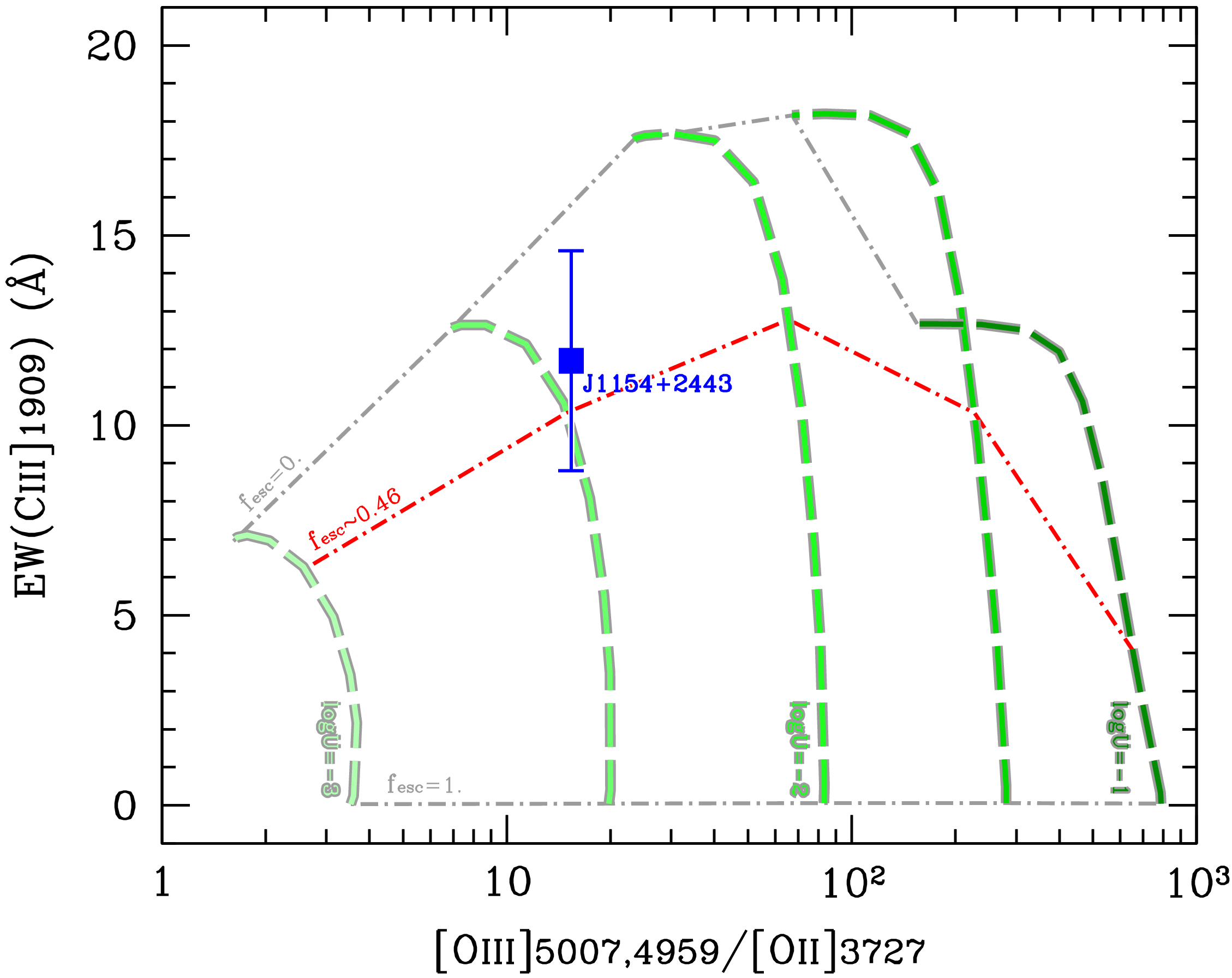}}
\caption{Predicted \Ciiiuv\ rest-frame equivalent width as a function of \Oiii/\Oii\ for a gas-phase metallicity $\oh = 7.65$ ($\sim 1/10$ solar), electron density $N_e=1000$ cm$^{-3}$, 
varying ionization parameters $\log U = -3, -2.5, -2, -1.5, -1$  (dashed colored lines from left to right)
and escape fractions from zero (top) approaching unity (bottom). The models assume a BPASS SED for a population of 1 Myr,
as in Fig.\ \ref{fig_c3}.
The observations of \source\ are well reproduced with a fairly high ionization parameter and with the
observed escape fraction $\fesc=46$ \%.}
\label{fig_cloudy}
\end{figure}

More quantitatively, we compare in Fig.\ \ref{fig_cloudy} the observed \Oiii/\Oii\ ratio, a measure of the
ionization parameter at known metallicity, and the \Ciii\ equivalent width to a grid of {\em Cloudy} \hii\ region models
for the metallicity of \source\ and  with varying degrees of Lyman continuum escape
\footnote{The models are an extension of those described in detail by \cite{Nakajima2018The-VIMOS-Ultra}, with
varying escape fractions computed as in \cite{Nakajima2014Ionization-stat}.}.
Compared to Fig.\ \ref{fig_c3} we have assumed a higher electron density ($N_e=1000$ cm$^{-3}$),
probably more appropriate for \source\ \citep[cf.][]{\paper}, although the uncertainty on $N_e$ remains large.
If correct, this leads to an increase of EW(\Ciii) by $\sim 5-20$\%.
The observations are well reproduced by models with a high ionization parameter ($\log U \sim -2.5$ to $-2$)
and escape fractions \fesc\ between 0 (ionization-bounded region) and $\sim 50$ \%, in agreement with
the measured value of \fesc. This comparison confirms that the 
%\LEt{If choosing to use an acronym, please spell it out in full the FIRST time it
%appears in the paper, followed by the abbreviation in parentheses, both in
%the abstract and again in the main text. After that, please only use the
%abbreviation. See A and A language guide Section 5.2.4 www.aanda.org/language-editing. Otherwise, please continue to use the full word.}
EW of UV metal lines cannot be used to
``single'' out, i.e.\ recognise, this galaxy as a strong LyC emitter, and to determine its LyC escape fraction.
More complex models will be needed in the future to account for geometrical constraints, such as
the existence of ionized holes and regions with neutral gas covering the UV source, as found for LyC leakers
by \cite{Gazagnes2018Neutral-gas-pro} and \cite{Chisholm2018Accurately-pred}.

Is the strength (EW) of other lines decreased due to a significant fraction of escaping ionizing photons ?
{\em Starburst99} models predict EW(\hb) consistently above 400 \AA\ at ages up to 3-4 Myr,
even higher up to 500 \AA\ at low metallicities \citep{SB99}.
The age of the UV-dominant population is constrained from optical continuum fits 
\citep[cf.][]{\paper} and from the UV lines (Schaerer et al., in prep.); it is of the order $\sim 1-3$ Myr.
The rest EW($\hb$)=160 \AA\ of \source\ is indeed significantly lower than these predictions,
compatible with a reduced equivalent width due to LyC escape. On the other hand the EW of optical lines
can also be reduced by the presence of an older underlying population, whose  observational features
may be difficult to find due to the strong emission lines and nebular continuum emission.
In any case, \cite{\paper} obtained a consistent model accounting for the observed Hydrogen line strengths
(fluxes and equivalent widths), an underlying population, and non-negligible LyC escape.

\cite{Zackrisson2013The-Spectral-Ev} proposed a method to identify galaxies with a high LyC escape fraction,
by combining rest-frame UV slope and EW(\hb) measurements to search for sources with reduced \hb\ 
equivalent widths. Clearly this method is complicated in practice, since it requires the determination of the
intrinsic UV slope, which depends on the a priori unknown amount of UV attenuation and the attenuation law
-- especially at high redshift \citep[see e.g.][]{Capak2015Galaxies-at-red,Popping2017Dissecting-the-,Narayanan2018The-IRX-beta-du} -- 
and is sensitive to several parameters including the metallicity, star-formation history, and age 
\citep{raiter2010,Binggeli2018Lyman-continuum}.
For \source\ we find a UV slope  $\beta \sim  -1.9 \pm 0.3$ from the STIS spectrum, 
considerably redder than the range of UV slopes predicted by \cite{Zackrisson2013The-Spectral-Ev},
and which is clearly affected by dust attenuation, as shown by \cite{\paper}. 
In short, we conclude that the method proposed by \cite{Zackrisson2013The-Spectral-Ev} 
is probably difficult to apply in practice to identify strong LyC leakers.

% % % % % % % % % % % % % % % % % % % % % % %
\subsection{No strong \Heiiuv\ emission}

The  \Heiiuv\ line is not detected in the present spectrum of \source. 
Our non-detection corresponds to a line ratio of \Ciiiuv/\Heiiuv$>9.6$ at $1 \sigma$, which, together with 
EW(\Ciii), places our source in the domain of star-forming galaxies \citep[cf.][]{Nakajima2018The-VIMOS-Ultra},  as expected.
However, the observations are not particularly deep, providing only a relatively weak limit of EW(\Heiiuv)$\la 3$ \AA. 
This does not exclude the presence of \heii\ emission (either stellar or nebular) in this galaxy, 
with a strength comparable to that often seen in low-metallicity systems \citep{Erb2010Physical-Condit,Senchyna2017Ultraviolet-spe}.
In the available optical spectrum, \Heiiopt\ is absent, with a relative  intensity $I(4686)/I(\hb) \la 0.02$ \citep{\paper}.
%Deeper spectra (optical or UV) are required to exclude the presence of \heii\ emission and better constrain 
%the hardness of the ionizing spectrum of this peculiar galaxy.

In any case, we note that similarly to \source, none of the 11 LyC leakers recently detected 
by our team show \Heiiopt\ emission (with comparable limits on $I(4686)/I(\hb)$), implying that the presence 
of a very hard ionizing spectrum is not a necessary criterion for LyC emission.

\begin{figure*}[ht]
{\centering
\includegraphics[width=9cm]{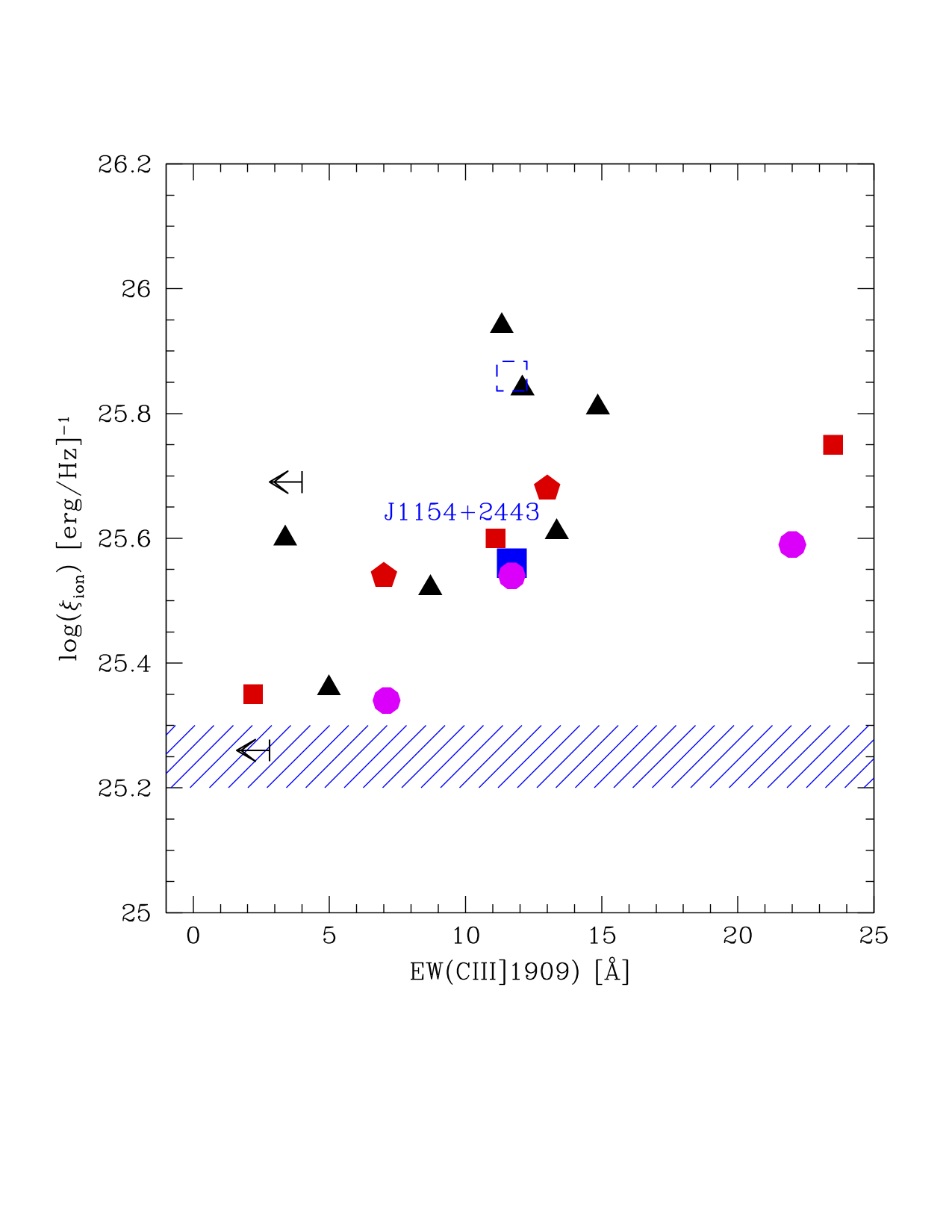}
\includegraphics[width=9cm]{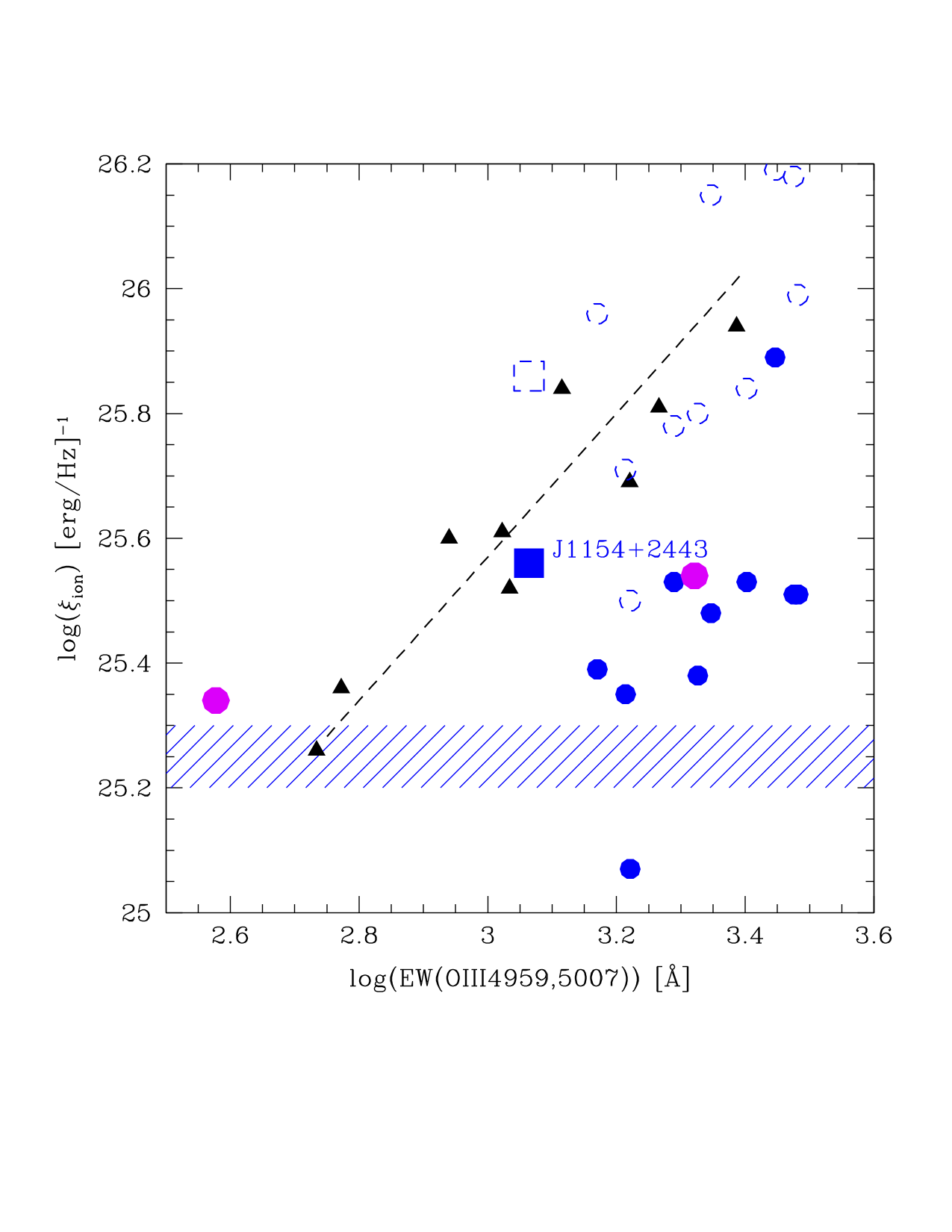}
%\vspace*{-2cm}
\caption{Attenuation-corrected (i.e.\ intrinsic) ionizing photon production, \chioncorr, as a function of the \Ciiiuv\ equivalent width ({\em left panel}) 
and the \Oiii\ EW ({\em right}).
\source\ is shown by the filled blue square; other LyC leakers as filled blue circles. Open symbols show \chion\ with no attenuation correction
for the UV luminosity.
Comparison samples are plotted using the same symbols as in Fig.\ \protect\ref{fig_c3}.
The blue shaded area denotes the range of \chioncorr\  often assumed as the ``standard'' value,
obtained for constant star formation over a long timescale and subsolar metallicities \citep[cf.][]{Robertson2013New-Constraints}.
}
}
\label{fig_chion}
\end{figure*}

% % % % % % % % % % % % % % % % % % % % % % %
\subsection{Ionizing photon production}

The ionizing photon production of galaxies is one of the fundamental measures of importance for
our understanding of cosmic reionization.
This quantity, that is,\ the rate of hydrogen ionizing photons produced in a galaxy per unit time, 
is very often expressed in units of UV luminosity, and commonly denoted as \chion. 
As in \cite{Schaerer2016The-ionizing-ph}, we have derived \chion\ for the newly discovered 
LyC emitters  of \cite{\paper,Izotov2018Low-redshift-Ly}, including \source,
from the Balmer recombination lines, which directly count the ionizing photon rate
and the COS UV spectra. Since some sources have a high LyC escape fraction we also
correct for this effect.
The bulk of the sources have $\log(\chioncorr)=25.3-25.6$ erg$^{-1}$ Hz, 
higher than the canonical values of  $\log(\chioncorr) \approx 25.2-25.3$ erg$^{-1}$ Hz
often adopted for high-$z$ studies \citep[e.g.][]{Robertson2013New-Constraints}.
Here \chioncorr\ denotes \chion\ normalised to the intrinsic UV luminosity at 1500 \AA\
rest-frame, after correction of dust attenuation. For \source,\ we obtain $\log(\chioncorr)=25.56$ erg$^{-1}$ Hz.

Determining the ionizing photon rate directly from UV spectral diagnostics would be very convenient,
especially for high-redshift studies.
For this purpose we plot \chioncorr\ as a function of the \Ciii\ equivalent width in Fig.\  \ref{fig_chion} (left).
\source\ has comparable \chioncorr\ and EW(\Ciii) to $z \sim 2-4$ \Ciii\ emitting galaxies observed in the VIMOS Ultradeep Survey
(VUDS) \citep[cf.\ sample B from][]{Nakajima2018The-VIMOS-Ultra}, faint narrow-band selected \lya\ emitters at $z=3.1$ 
\citep{Nakajima2018The-Mean-Ultrav}, the strong emission line galaxy of \cite{Berg2018A-Window-On-The},
and to some low-$z$ galaxies studied  by \cite{Senchyna2017Ultraviolet-spe}.
Clearly a tendency of increasing \chioncorr\ with increasing \Ciiiuv\ equivalent width is found
\citep[cf.\ also][]{Chevallard2018Physical-proper}. 
Such a trend is not unexpected as \chioncorr\ increases for young and metal-poor stellar populations
\citep[see e.g.][]{schaerer2003,raiter2010}, which are conditions leading also to stronger \Ciii\ emission
\citep[cf.][]{Jaskot2016Photoionization,Nakajima2018The-VIMOS-Ultra}.
On the other hand, the \Ciii\ line strength depends also on the C/O abundance ratio in the gas phase, which may vary from
object to object, and the EW is predicted to decrease again for metallicities below $(0.1-0.2)$ times solar,
as shown by \cite{Nakajima2018The-VIMOS-Ultra}. This predicted ``downturn'' has not yet been seen clearly
(cf.\ Fig \ref{fig_c3}).
In any case,  evolutionary synthesis models predict a maximum $\log(\chion)=25.8$ erg$^{-1}$ Hz at the youngest ages
and for low metallicities ($Z \ga 1/100$ \zsun); for PopIII this can go up to $\log(\chion)=26.2$ erg$^{-1}$ Hz, depending
on the 
%\LEt{Please spell out all acronyms the first time they
%appear in the paper, followed by the abbreviation in parentheses, both in
%the abstract and again in the main text. After that, please only use the
%abbreviation. See A and A language guide Section 5.2.4 www.aanda.org/language-editing} 
stellar initial mass function \citep[see][]{raiter2010}.

% BERG z=1.844 source, 12+;og(O/H)=***
% rest-frame EW(CIII)=17.3 Ang

% Flam=0.6e-17 erg/cm2/s/Ang at observed 4266 Ang --> mag_AB=22.49
%       737 cosmology: M_AB=-22.12 OK --> intrinsin M_1500=-19.02 After lensing correction with mu=17.3
%     d_L= 14065.0 Mpc --> DM=44.606 
% Q_ion=7.3e53 /s after extinction and lensing correction with mu=17.3 (sect 8.2)
% --> chi_ion=25.6   using M_1500 without attenuation correction 
% --> chi_ion=25.54 using M_1500 and factor 1.15  attenuation correction  (their Table 5 for CIV)

% ERB+2016 z=2.3 source BX418, 12+log(O/H)=7.9+-0.2
% rest-frame EW(CIII)=7.1+-0.4 Ang
%
% SFR(Ha)=15 no ext correction (from NIRSPEC spectrum) --> L(Ha)=1.90e42 erg/s (K98) --> Q(H)=1.40e54 /s for std Te
% SFR(UV)=9 --> L_nu=6.43e28
% --> log(chi_ion) = 25.34
%

Other indirect estimates of the ionizing photon production have been proposed, for example\ using the  rest-frame EW of the optical \Oiii\ lines
\citep{Chevallard2018Physical-proper}. 
Figure \ref{fig_chion} (right) shows \chioncorr\ as a function of EW( \oiii) 
and the proposed correlation of these quantities from \cite{Chevallard2018Physical-proper}.
% (black dashed line).
Although \source\ is in agreement with this relation, the vast majority of the known low redshift LyC emitters, shown by the filled blue symbols,  
clearly do not 
follow this correlation, except if the ionizing photon production is normalised by the observed (not the intrinsic) UV luminosity
(as shown by the dashed open circles). 
In any case the physical origin of such a correlation is more difficult to explain, especially since in general EWs of optical emission lines 
are subject to ``dilution'' from older stellar populations, that is,\  they are dependent on the past star-formation history.

%%%%%%%%%%%%%%%%%%%%%%%%%%%%%%%%%%%%%%%%%%%%%%%%%%%%%%%%%%%%%%%%%%%%%%
\section{Conclusion}
\label{s_conclude}

We have obtained the first complete UV spectrum of a low-redshift galaxy with strong LyC emission, \source,
which was recently discovered by \cite{\paper}, and which shows a very high escape fraction of LyC radiation.
This galaxy is a low-metallicity ($\oh = 7.65$), low-mass ($\log M_\star \sim 8.2$ \msun), compact star-forming galaxy, which had been selected 
for its
%\LEt{Please check that I have retained your intended meaning.} 
very high ratio of the optical lines \oiiil / \Oii $=11.5$.
The observations with STIS on  board HST, covering the spectral range $\sim 1200-2200$ \AA\ restframe, show 
strong \lya\ and \Ciiiuv\ emission, as well as the presence of \Oiiiuv. 

We find a C/O abundance of $\log({\rm C/O})\sim -0.9$; that is, low, but comparable to other C/O measurements at low metallicity.
\source\ shows a very strong \Ciiiuv\ emission line with an equivalent width EW(\Ciii)$=11.7 \pm2.9$ \AA,
comparable to some other low-redshift sources of similarly low metallicity (see Fig.\ \ref{fig_c3}).
The high EW(\Ciii) shows that {\em strong} LyC emitters do not necessarily have weak \Ciii\ emission, as predicted by
some models \citep[cf.][]{Jaskot2016Photoionization}.
Simple photoionization models can explain the main observational features and the UV spectrum of \source,
but are not able distinguish between no LyC escape and the observed \fesc=$46$ \%.

The intrinsic ionizing photon production efficiency of \source\ is  $\log(\chioncorr)=25.56$ erg$^{-1}$ Hz, comparable to that
of the other recently discovered $z \sim 0.3-0.4$ LyC emitters \citep[see][]{Schaerer2016The-ionizing-ph}, and higher than the canonical value for \chioncorr\ by a factor of approximately two \citep[cf.][]{Robertson2013New-Constraints}.
With other data from the literature, we find a trend of 
increasing \chioncorr\ with increasing \Ciiiuv\ equivalent width (Fig.\ \ref{fig_chion} left), which can be understood
by a combination of decreasing stellar population age and metallicity.
If confirmed with larger samples, such a relation would be useful for studies of high-$z$ galaxies, which rely on rest-UV spectra.
The majority of the $z \sim 0.3-0.4$ leakers discovered with HST/COS observations do not follow the  correlation
between  \chioncorr\ and EW(\Oiii) proposed by \cite{Chevallard2018Physical-proper}.

%%%%%%%%%%%%%%%%%%%%%%%%%%%%%%%%%%%%%%%%%%%%%%%%%%%%%%%%%%%%%%%%%%%%%%%%%%%%%%%%%
\begin{acknowledgements}
DS wishes to dedicate this publication to the memory of Nolan Walborn, passed away in february 2018,
who was an exceptional spectroscopist,  inspiring colleague, and a very humorous and open-minded person.

We thank Danielle Berg for communications on low-redshift galaxy samples.
This work is based on HST mid-cycle observations (GO 15433, PI Schaerer), for which we thank the HST staff for their help.
Y.I. acknowledges support from the National Academy of Sciences of Ukraine (Project No. 0116U003191) and 
by its Program of Fundamental Research of the Department of Physics and Astronomy (Project No. 0117U000240).
K.N. acknowledges a JSPS Research Fellowship for Young Scientists, A.V.  funding from the ERC-stg-757258 grant  TRIPLE,
and T.X.T.  support from grant HST-GO-15433.002-A.

\end{acknowledgements}
%%%%%%%%%%%%%%%%%%%%%%%%%%%%%%%%%%%%%%%%%%%%%%%%%%%%%%%%%%%%%%%%%%%%%%%%%%%%%%%%%
\bibliographystyle{aa}
%\bibliography{references}
\bibliography{merge_misc_highz_literature}

\end{document}